\documentclass[prb,a4paper,twocolumn,showpacs,floatfix,superscriptaddress]{revtex4}
\usepackage{graphicx}
\usepackage{amsmath,amssymb}
\usepackage{color}

\begin{document}

\title{Analytic Determination of the Critical Coupling for Oscillators in a Ring}

\author{Hassan F. El-Nashar}

\affiliation{Department of Physics, Faculty of Science, Ain Shams University, 11566 Cairo, Egypt}

\affiliation{Department of Physics, Faculty of Education, King Saud University, P.O.~Box 21034, 11942 Alkharj, K.S.A}

\author{Hilda A. Cerdeira}

\affiliation{Instituto de F\'{\i}sica Te\'orica, Universidade Estadual Paulista, R. Pamplona 145, 01405-000 S\~ao Paulo, Brazil}

\affiliation{Instituto de F\'{\i}sica, Universidade de S\~ao Paulo, R. do Mat\~ao, Travessa R. 187, 05508-090 S\~ao Paulo, Brazil}

\date{\today}

\begin{abstract}
We study a model of coupled oscillators with bidirectional first nearest neighbours coupling with periodic boundary conditions. We show that a stable
phase-locked solution is decided by the oscillators at the borders between the major clusters, which merge to form a larger one of all oscillators at
the stage of complete synchronization. We are able to locate these four oscillators as well as the size of major clusters in the vicinity of the stage
of full synchronization which we show to depend only on the set of initial frequencies. Using the method presented here, we are able to obtain an
analytic form of the critical coupling, at which the complete synchronization state occurs.
\end{abstract}


\maketitle
{\bfseries Weakly coupled oscillators in the chaotic state have been known
to represent many physical systems, as well as chemical,
biological, neurological and so on. These systems synchronize
in frequency under the influence of coupling. Knowing
beforehand the value of the coupling constant and the dynamical
behavior of the individual oscillators for complete
synchronization to occur is an important source of information
for real applications. This paper is a continuation of previous
theoretical results for these systems. Here, we derive
relationships that allow us to determine the oscillators which
first lock in phase and drag the whole system into the
synchronized state as well as the size of the two existing
clusters before the transition.}

\section{Introduction}
In recent years we have seen oscillators coupled through nearest neighbors interactions to be used to understand the behavior of systems in physics,
chemistry, biology, neurology as well as other disciplines, to model several phenomena such as: Josephson junction arrays, multimode lasers, vortex
dynamics in fluids, biological information processes, neurodynamics \cite{1,2,14,suz}. These systems have been observed to synchronize themselves to a
common frequency, when the coupling strength between these oscillators is increased \cite{suz,hak,15}. In spite of the diversity of the dynamics, the
synchronization features of many of the above mentioned systems might be described using a simple model of weakly coupled phase oscillators such as the
Kuramoto model \cite{15,17,suz,hak}, as well as its variations to adapt it for finite range interactions which are more realistic to mimic many
physical systems. Difficulties arise since finite range coupled systems are difficult to analyze and to solve analytically. In spite of that, in order
to figure out the collective phenomena when finite range interactions are considered, it is of fundamental importance to study and to understand the
nearest neighbour interactions, which is the simplest form of the local interactions. In this context, a simplified version of the Kuramoto model with
nearest neighbour coupling in a ring topology, which we shall refer to as \emph{locally coupled Kuramoto model} (LCKM), is a good candidate to describe
the dynamics of coupled systems with local interactions. Several reports exist where the LCKM has been used to represent the dynamics of a variety of
systems, such as Josephson junctions, coupled lasers, neurons, chains with disorders, multi-cellular systems in biology and in communication systems
\cite{hak,17,arx,wie1}. It has also been shown that the equations of the resistively shunted junction which describe a ladder array of overdamped,
critical-current disordered Josephson junctions that are current biased along the rungs of the ladder can be expressed by a LCKM  \cite{16}. For
nearest neighbours coupled R\"ossler oscillators the phase synchronization can be described by the LCKM \cite{22}, as well as locally coupled lasers
\cite{wie2,wie3}, where local interactions are dominant. LCKM can also be used to model the occurrence of travelling waves in neurons \cite{hak,suz}.
In communication systems, unidirectionally coupled Kuramoto model can be used to describe an antenna array~\cite{iop}. Such unidirectionally coupled
Kuramoto models can be considered as a special case of the LCKM and it often mimics the same behaviour. Therefore, LCKM can provide a way to understand
phase synchronization in coupled systems in general.

While in the Kuramoto model for long range interactions one has to rely on average quantities, in a mean field approximation or by means of an order
parameter, etc., in the local model it is necessary to study the behaviour of individual oscillators in order to understand the collective dynamics.
Therefore, due to the difficulty in applying standard techniques of statistical mechanics, one should look for a simple approach to understand the
coupled system with local interactions by means of numerical study of a temporal behaviour of the individual oscillators. Such analysis is necessary in
order to obtain a close picture of the effect of the local interactions on synchronization. In this case, numerical investigations can assist to figure
out the mechanism of interactions at the stage of complete synchronization which in turn helps to get an analytic solution. Earlier studies on the LCKM
show several interesting features including tree structures with synchronized clusters, phase slips, bursting behaviour and saddle node bifurcation and
so on \cite{stro1,18}. It has also been shown that neighbouring elements share dominating frequencies in their time spectra, and that this feature
plays an important role in the dynamics of formation of clusters in the local model \cite{19}; that the order parameter, which measures the evolution
of the phases of the nearest neighbour oscillators, becomes maximum at the partial synchronization points inside the tree of synchronization \cite{20}
and a scheme has been developed based on the method of Lagrange multipliers to estimate the critical coupling strength for complete synchronization in
the local Kuramoto model with different boundary conditions \cite{21}.

Very recently, we identified two oscillators which are responsible for dragging the system into full synchronization \cite{ms-last}, and the difference
in phase for this pair is $\pm\pi/2$. In this work we develop a method to obtain an analytic solution for the value of the critical coupling at which
full synchronization occurs, once a set of initial conditions for the frequencies of the $N$ oscillators is assigned. This method will allow us not
only to calculate the analytic form of the critical coupling but also to determine the number of oscillators at the major clusters in the vicinity of
the critical coupling as well as to determine which is the pair of oscillators that has a phase difference $\pm\pi/2$ at the stage of full
synchronization.

This paper is organized as follows. In Sec.~\ref{ring} we investigate the LCKM where periodic boundary conditions are used. We derive an analytic form
for the critical coupling at the stage of complete synchronization as well and determine the number of oscillators at each cluster in the vicinity of
the critical coupling. Finally, in Sec.~\ref{finalwords} we give a conclusion which is based on a summary of the results.

\section{oscillators in a ring}
\label{ring} The local model of nearest neighbour interactions, or LCKM can be considered as a diffusive version of the Kuramoto model, and it is
expressed as \cite{18,19,20,21,ms-last}:
\begin{align}
\dot{\theta}_i= &\ \omega_i+K[\sin(\phi_i) - \sin(\phi_{i-1})], \label{eqring}
\end{align}
with periodic boundary conditions $\theta_{i+N}=\theta_i$ and for $i=1,2,...,N$. The set of the initial values of frequencies $\{\omega_i\}$ are the
natural frequencies which are taken from a Gaussian distribution and $K$ is the coupling strength. The phase difference is defined as
$\phi_i=\theta_{i+1}-\theta_i$ for $i=1,2,3,...,N$. These nonidentical oscillators (\ref{eqring}) cluster in time averaged frequency, until they
completely synchronize to a common value given by the average frequency $\omega_0 = \frac{1}{N}\sum_{i=1}^{N}\omega_i$, at a critical coupling $K_c$.
\begin{figure}[!ht]
\centering\includegraphics[width=\linewidth,clip]{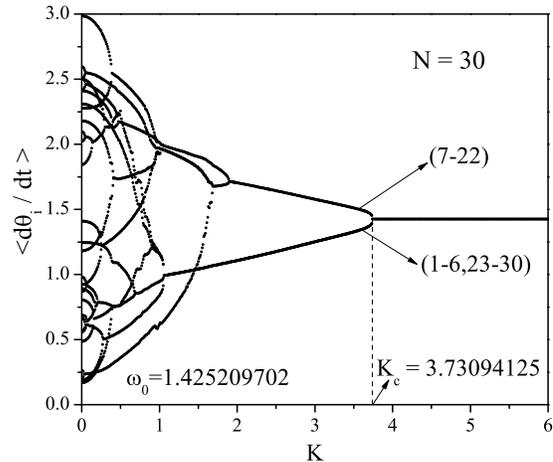} \caption{Synchronization tree for a system of $30$ oscillators, with detailed composition of
each cluster before full synchronization.} \label{figure1}
\end{figure}
At $K \ge K_c$ the phases and the frequencies are time independent and all the oscillators remain synchronized. In Fig.1 we show the synchronization
tree for a periodic system with $N=30$ oscillators, where the elements which compose each one of the major clusters are indicated in each branch. These
clusters merge into one at $K_c$ where all oscillators have the same frequency. The major clusters just before $K_c$ contain $N_1$ and $N_2$
oscillators, where $N=N_1+N_2$. It is not necessary for these clusters to have the same numbers of oscillators. At the vicinity of $K_c$, major
clusters of successive oscillators have sets of nearest neighbours at the borders. An interesting fact emerges: the phase-locked solution is always
valid for one and only one phase difference, and this phase difference is between two oscillators at the border of the clusters \cite{ms-last}. Thus,
for these two neighbouring oscillators, the equation for the phase difference is:
\begin{align}
\dot \phi_n=\Delta_n - 2K \sin(\phi_n) + K \sin(\phi_{n-1}) + K \sin(\phi_{n+1}), \label{pheql}
\end{align}
where $\Delta_n=\omega_{n+1}-\omega_n$. Equation (\ref{pheql}) at $K_c$ has $\dot\phi_n=0$, and hence $\dot\theta_n=\dot\theta_{n+1}=\omega_0$. It has
been found that the phase-locked solution is satisfied when $\phi_n=\pi/2$ for the case of $\omega_{n+1}>\omega_n$ and $\phi_n=-\pi/2$ for the reverse.
In addition the phase-locked solution exists at \cite{ms-last} $X_n = |\frac{\Delta_n}{K_c} + \left[ \sin(\phi_{n-1})+\sin(\phi_{n-1}) \right]|=2$. It
is already well known that in the vicinity of $K_c$, phases as well as frequencies present the phenomenon of phase slip, that is, they remain constant
for a given period of time $T$ and then they jump, followed by a another period of constant value, a jump and so on. During this period $T$ equation
(\ref{pheql}) can be solved analytically to give quantities $\phi_n$ and $\dot\phi_n$. Due to the diffusive character of the LCKM, all other quantities
$\phi's$ and $\dot\phi's$ of other oscillators relate to $\phi_n$ and $\dot\phi_n$ (see details in reference \cite{ms-last}), and will in turn present
the same phenomena, which has also been seen numerically \cite{18,22}. Thus, if one can develop a method to allocate the two oscillators which will
have the phase difference $\phi_n=\pm\pi/2$, then it will be possible to determine the value $X_n$ and to obtain the critical coupling at which a
complete synchronization occurs ($K_c$). A difficulty arises due to the determination of the values of $\sin(\phi_{n+1})$ and $\sin(\phi_{n-1})$ in
addition to the topology of a ring which is an endless system. Therefore, there is no direct method to specify the two oscillators $n$ and $n+1$, which
have phase difference that would satisfy the phase-lock condition.

\begin{figure}[!ht]
\centering\includegraphics[width=\linewidth,clip]{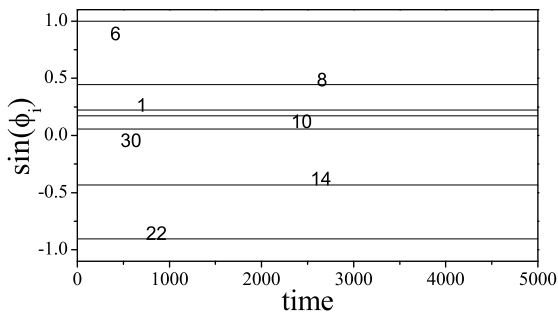} \caption{Selected Values of $\sin \phi_i$ at $K = K_c$ for the system of $30$ oscillators of
Fig.1.} \label{figure2}
\end{figure}
We can take advantage that there are four oscillators, now labeled $l$, $l+1$, $m$ and $m+1$, at the borders of the major clusters in the vicinity of
$K_c$ from which only one pair will have a phase difference corresponding to the phase-locked solution; i.e, $|\sin(\phi_n)|=1$. As shown in Fig. 2,
the values of $\sin(\phi_l)$ and $\sin(\phi_m)$ are always the maximum and minimum of the $\sin(\phi_i)$'s for all phase differences. From these two
phase differences $\phi_l$ and $\phi_m$, one of them has a value $\pm\pi/2$, while the other would be close $\mp\pi/2$, getting closer as $N$
increases. This fact has been verified numerically for several realizations of $N$ and $\{\omega_i\}$'s. If we start adding equations of the systems in
a ring (\ref{eqring}) (adding elements) in a similar way as in reference \cite{stro1}, we generate a sequence $Z_i$, with $Z_i=i
\omega_0-\sum_{j=1}^{i}\omega_j$ for $i=2,3,...,N-1$. After a detailed study of existing correlations, we arrive to a criteria to determine the four
oscillators at the borders of the major clusters in the vicinity of $K_c$. We find that the maximum value of $Z_i$ refers always to the oscillators at
one border while the minimum of $Z_i$ points to the oscillators at the other border of the given cluster. One of them is always positive and the other
is negative. The sign of $Z_i$ depends on the values of $\Delta_i=\omega_{i+1}-\omega_i$: for $max(Z_i)$, $\Delta_i>0$, while $min(Z_i)$ corresponds to
$\Delta_i<0$. A thorough study also shows that the value of $\sin(\phi_i)>0$ corresponds to $max(Z_i)$ while the $\sin(\phi_i)<0$ corresponds to
$min(Z_i)$. Thus we can calculate all the $Z_i's$ and assign the maximum and minimum values which refer to the integers $l$ and $m$. However, we have
not yet resolved which one refers to the two oscillators with the phase difference equals $\pm\pi/2$.

Since now we know how to point to the four oscillators $l$, $l+1$, $m$ and $m+1$ at the borders of major clusters at $K_c$, it is possible to obtain an
analytic form for $K_c$, within reasonable accuracy. These two quantities $Z_l$ and $Z_m$ are related to both values of $\phi_l$ and $\phi_m$ for a
system of coupled oscillators in a ring (\ref{eqring}) such that:
\begin{align}
K_c\left[\sin(\phi_l) - \sin(\phi_N)\right] = Z_l \\ K_c\left[\sin(\phi_m) - \sin(\phi_N)\right]= Z_m. \label{eqz}
\end{align}
The value of $\sin(\phi_N)$ presents a difficulty in determining which one of the phase differences $\phi_l$ or $\phi_m$ corresponds to $|\pi/2|$,
since the phase-lock condition at the critical coupling will be \cite{stro1} either $K_c=Z_l+K_c\sin(\phi_N)$ or $K_c=Z_m+K_c\sin(\phi_N)$. We can then
rely on numerical findings and characteristics of $\phi_l$ and $\phi_m$ to use equations (3) and (4) to obtain an approximate analytic expression for
$K_c$. A detailed numerical investigation of $\sin(\phi_N)$ shows that it is always small, in comparison to both values $Z_l$ and $Z_m$, and that $K_c$
depends mainly only on these two quantities $Z_l$ and $Z_m$. Taking this fact into consideration and since both $\sin(\phi_l)$ and $\sin(\phi_m)$ are
always opposite in sign as well as the two quantities $Z_l$ and $Z_m$, we use equations (3) and (4) to obtain
\begin{align}
K_c = \frac{|Z_l|}{2} + \frac{|Z_m|}{2} + \epsilon, \label{eqkc2}
\end{align}
where $\epsilon$ depends on the difference between both $|\sin(\phi_l)|$ and $|\sin(\phi_m)|$. Helping ourselves by numerical studies we find that
$\epsilon$ depends on the quantities $\Delta_l$ and $\Delta_m$ and it can can be written as $\epsilon\approx\frac{|\Delta_l|+ |\Delta_m| - |\Delta_l +
\Delta_m|}{16}$. Thus we obtain an approximate expression for $K_c$, which we call $K_c^a$, and is given by:
\begin{align}
K_c^a \approx &\ \frac{|Z_l| + |Z_m|}{2} +  \frac{|\Delta_l|+|\Delta_m| - |\Delta_l+\Delta_m|}{16}. \label{eqkc3}
\end{align}
\begin{figure}[!ht]
\centering\includegraphics[width=\linewidth,clip]{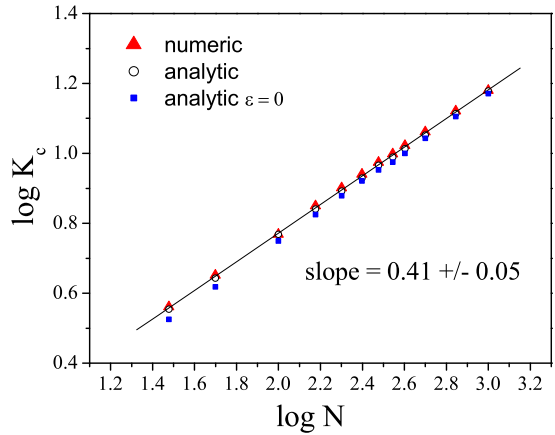} \caption{(Color online) log-log plot of $K_c$ versus $N$ from numerical simulation of system
(1) (triangles) and analytic calculation using equation (6) (circles) and equation (5) with $\epsilon=0$ (squares).} \label{figure3}
\end{figure}
Fig.3 summarizes numerical investigations of the determinations of $K_c$. We plot log$K_c$ versus log$N$ from numerical simulations of equation
(\ref{eqring}) (triangles) and from the analytic results given by (\ref{eqkc3}), first by considering $\epsilon$ going to zero (squares) and then
taking its complete dependence on both $\Delta_l$ and $\Delta_m$ (circles). The validity of equation (6) is clearly shown for values of $N$ ranging
from $30$ to $1000$. The dependence of $K_c$ on both $|Z_l|$ and $|Z_m|$ as in equation (\ref{eqkc3}) and as $N$ increases becomes clear.  It can also
be inferred that the term which depends on $\epsilon$ becomes negligible. This is due to the fact that as $N$ increases the oscillators of indexes $l$
and $l+1$ are becoming closer in frequencies to each other as well as the two oscillators of indexes $m$ and $m+1$. We also observe that $K_c$ grows as
$\sim O\sqrt{N}$, in the same limit as found by Strogatz and Mirollo \cite{stro1} (for details, see explanation in this reference).

Summarizing, if one knows the set of initial frequencies $\{\omega_i\}$, it is possible to point at the four oscillators at the borders of the major
clusters just below $K_c$ and then the calculation of $K_c$ is performed using equation (6) (thus obtaining $K_c^a$), without the need of computer
simulation of system (\ref{eqring}), just using the values of $Z_l$ and $Z_m$. If we are interested in determining which phase difference will have a
phase-lock condition $\pm\pi/2$, we use the fact that $\sin(\phi_l)$ and $\sin(\phi_m)$ have opposite signs as well as they are maximum and minimum
among all values of sine of the phase differences. The sign of the quantity $\sin(\phi_N)$ has the same sign of the quantity $x_1=-(Z_l+Z_m)/2$, which
is taken from the sum of equations (3) and (4) (eliminating for a moment the small difference between $\sin(\phi_l)$ and $\sin(\phi_m)$). Depending on
the signs of $Z_l$ and $Z_m$, we know the signs of $\sin(\phi_l)$ and $\sin(\phi_m)$, and hence the sign of $\sin(\phi_N)$. Therefore, we count two
quantities $x_2=\pm K_c^a-Z_l$ and $x_3=\pm K_c^a-Z_m$, positive sign for $Z>0$ and negative sign for the reverse. Three cases will exist: first from
the quantities $x_2$ and $x_3$, one is positive and the other is negative. Thus depending on the sign of $x_1$ we choose either $x_2$ or $x_3$ to be
$K_c^a\sin(\phi_N)$. Second $x_2$ and $x_3$ have the same signs, then we check the minimum between $|x_1-x_2|$ and $|x_1-x_3|$ and depending on which
one is the minimum, we take either $x_2$ or $x_3$ to be $K_c^a\sin(\phi_N)$. Third $|x_1-x_2|=|x_1-x_3|$, then we take the minimum outcome of $x_2$ and
$x_3$. Now we know the value of $K_c^a\sin(\phi_N)$ and its sign. Therefore, we know which equation from (3) and (4) will be used to give $K_c^a$. Thus
we specify which phase difference of index $l$ or $m$ would have $\pm\pi/2$. We tested this method on the simulations we have done and it matches the
outcome of the numerical simulations.

The number of oscillators in each cluster at the vicinity of $K_c$ can be determined, once we assigned the indexes $l$, $l+1$, $m$ and $m+1$, which, we
remind the reader, are obtained from $Z_l$ and $Z_m$, maximum and minimum values of the sequence $Z_i$. The size of one cluster of $N_1$ oscillators is
determined by counting the difference $N_1 = (m+1)-l$ and the size of the other cluster is determined as $N_2=N-N_1$. Similarly to the calculation of
$Z_l$ and $Z_m$, we can determine other two quantities which are $Y_1=N_1\omega_0 - \sum_{i=l+1}^{m}\omega_i$ and $Y_2=N_2\omega_0 -
\sum_{i=m+1}^{l}\omega_i$ taking into consideration the periodic boundary conditions. It is found that $|Y_1|=|Y_2|$. These quantities are related to
$Z_l$ and $Z_m$ by $Y_1=Z_m-Z_l=-Y_2$. It is easy to show that $K_c=\frac{|Y_1|}{2} + \epsilon =\frac{|Y_2|}{2} + \epsilon$. The two quantities $Y_1$
and $Y_2$ provide a criteria to understand synchronization-desynchronization at $K_c$. If one arrives from above $K_c$ where all oscillators are
synchronized and have the same value of frequency, at $K_c$ the oscillators split into two groups of $N_1$ and $N_2$, at $K_c$, depending on these two
quantities $Y_1$ and $Y_2$, where $|Y_1|=|Y_2|$. It is not necessary for $N_1$ to be equal to $N_2$. Both quantities $Y_1$ and $Y_2$ have opposite
signs since they refer to two groups of oscillators (two clusters) one of them rotates with average frequency over $\omega_0$ and the other has an
average frequency lower than $\omega_0$.

Comparing our findings of $K_c$ with the work of Daniels et al. \cite{16}, our method has the advantage of finding the value of $K_c$ without
performing numerical simulations once we know the set of initial frequencies $\{\omega_i\}$. In addition we get the condition of
synchronization-desynchronization at $K_c$ and obtain the number of oscillators in each branch in the vicinity of $K_c$.

\section{conclusion}
\label{finalwords} We have analyzed the conditions of the phase differences for the onset of complete synchronization at the critical coupling strength
in a Kuramoto-like model with nearest neighbour coupling with periodic boundary conditions. Such analysis allows us to determine the four oscillators
located at the borders of the major clusters (formed by successive oscillator) which will meet at the critical coupling to form one cluster of all
synchronized oscillators. With the help of these findings we derive the analytic expression for the critical coupling when all oscillators will have
the same frequency and, phase differences and instantaneous velocities become time independent. In addition, we are able to determine which is the
phase difference, that will have a phase-lock solution $\pm\pi/2$. From the derivation we also extract the size of the clusters before complete
synchronization. The analytic form of $K_c$ depends only on the initial frequencies, through the quantities $Z_l$ and $Z_m$, where the indexes $l$ and
$m$ correspond to the borders of the clusters. The quantities $Z_l$ and $Z_m$ correspond to the maximum and minimum values of the sequence $Z_i$. These
quantities in fact are related to the statistics of the distribution of the set of initial frequencies $\{\omega_i\}$, when this sample is obtained
from a Gaussian distribution, as shown by Strogatz and Mirollo \cite{stro1}. A detailed study within this context could shine light on the behavior of
$K_c$ as $N \rightarrow \infty$, not just for the case of a Gaussian distribution, but for others. This investigation plus extension of the method to
study cluster formation inside the tree will be topics of further analysis. The advantages of the study presented here is that we can determine the
value of the coupling constant that will synchronize the system of coupled oscillators without carrying out numerical simulation as well as to
determine the sizes of the clusters just before this happens. Generalization of these results to different couplings and boundaries are under
investigation and will be presented elsewhere.

\acknowledgments

HFE thanks the Abus Salam ICTP, Trieste, Italy, for hospitality during part of this work.

\end{document}